# Mapping Complex Mode Volumes with Cavity Perturbation Theory

K. G. Cognée,[1,2] W. Yan,[1] F. La China,[3] D. Balestri,[3] F. Intonti,[3] M. Gurioli,[3] A. F. Koenderink,[2] and P. Lalanne[1]*

[1]LP2N, Institut d'Optique, CNRS, Univ. Bordeaux, Talence, 33400, France
[2]Center for Nanophotonics, AMOLF, Science Park 104, 1098XG, Amsterdam The Netherlands
[3]LENS, University of Florence, Sesto Fiorentino, 50019, Italy
E-mail: philippe.lalanne@institutoptique.fr

**Abstract**
Microcavities and nanoresonators are characterized by their quality factors *Q* and mode volumes *V*. While Q is unambiguously defined, there are still questions on V and in particular on its complex-valued character, whose imaginary part is linked to the non-Hermitian nature of open systems. Helped by cavity perturbation theory and near field experimental data, we clarify the physics captured by the imaginary part of V and show how a mapping of the spatial distribution of both the real and imaginary parts can be directly inferred from perturbation measurements. This result shows that the mathematically abstract complex mode volume in fact is directly observable.

Predicting how the presence of a tiny foreign object near a resonant optical cavity perturbs the optical response is a classical problem in electromagnetics, with important implications spanning from the radio-frequency domain to present-day nano-optics. The perturbation results in a modification $\Delta\widetilde{\omega}$ of the initial complex resonance frequency $\widetilde{\omega} \equiv \omega_0 + i\gamma_0/2$ of the unperturbed cavity mode, $\text{Re}(\Delta\widetilde{\omega})$ and $\text{Im}(\Delta\widetilde{\omega})$ respectively representing the frequency shift and linewidth change. For a tiny perturbation quantified by a dipolar polarizability $\alpha$ (assumed small and isotropic) and placed at $\mathbf{r}_0$, $\Delta\widetilde{\omega}$ usually reads as

$$\frac{\Delta\widetilde{\omega}}{\widetilde{\omega}} \approx \frac{-\alpha\,\varepsilon(\mathbf{r}_0)|\widetilde{\mathbf{E}}(\mathbf{r}_0)|^2}{\iiint\left[\varepsilon|\widetilde{\mathbf{E}}|^2 + \mu_0|\widetilde{\mathbf{H}}|^2\right]d^3r} \equiv \frac{-\alpha}{2V(\mathbf{r}_0)}, \qquad (1)$$

where $\varepsilon$ is the permittivity of the unperturbed cavity, $\varepsilon_0$ and $\mu_0$ are vacuum permittivity and permeability, and $\widetilde{\mathbf{E}}$ and $\widetilde{\mathbf{H}}$ are the unperturbed-cavity-mode electric and magnetic fields. The seminal Eq. (1) has been initially proposed by Bethe-Schwinger in optics [1], and Waldron in the radio frequency domain [2-3], and has been used in similar variants until recently [3-7]. For convenience, we have introduced the mode volume $V$, the classical *real* quantity used throughout in quantum electrodynamics [8-9] that gauges the coupling of an emitting dipole with the cavity mode. $V$ is usually defined for dipoles placed at the field-intensity maximum, where the coupling is also maximum. For convenience, we rather consider a spatially-dependent mode volume to directly take into account the dependence of $\Delta\widetilde{\omega}$ with the perturber position. Equation (1) has the merit of being intuitive and easy to evaluate, since $\Delta\widetilde{\omega}$ solely depends on the unperturbed mode. It has been widely used for determining the dielectric and magnetic parameters of materials or testing the functionalities of microwave circuit components [3], and in the optical domain, to detect [10-11] or trap [7] nanoparticles, tune the resonance of photonic-crystal (PhC) cavities [4,12-17], analyze the impact of fabrication imperfections on these cavities [5], or study magnetic-like light-matter interactions [6,18]. Remarkably, Equation (1) cannot accurately predict perturbation-induced changes of the quality factor, $Q = -\frac{\text{Re}(\widetilde{\omega})}{2\,\text{Im}(\widetilde{\omega})}$. In particular, it predicts that changes in cavity loss rate follow the exact same spatial dependence as changes in the real frequency, with the sign of the polarizability setting the sign of the change in loss rate. This issue is known since the very beginning of perturbation theory and is sometimes accounted for by appending an additional flux-like term to Eq. (1) [3], even in recent works [17]. This term unfortunately requires solving the perturbed problem.

With the recent advent of theoretical results on the normalization of leaky resonator modes [8,19-20], it becomes evident that cavity perturbation theory cannot rely a normalization based on energy but on quasinormal-mode (QNM) formalism to account for the non-Hermitian character of the problem. Thus it has been proposed recently that Eq. (1) is conveniently replaced by

$$\frac{\Delta\widetilde{\omega}}{\widetilde{\omega}} \approx \frac{-\alpha\,\varepsilon(\mathbf{r_0})\widetilde{\mathbf{E}}^2}{\iiint[\varepsilon\widetilde{\mathbf{E}}^2 - \mu_0\widetilde{\mathbf{H}}^2]d^3r} \equiv \frac{-\alpha}{2\widetilde{V}(\mathbf{r_0})}. \qquad (2)$$

The sole difference between Eqs. (1) and (2) is the replacement of the *real* modal volume $V$ by a *complex* one $\widetilde{V}$, which is calculated from the QNM field distribution ($\widetilde{\mathbf{E}}$, $\widetilde{\mathbf{H}}$), see Section 1 in SM for a more formal comparison. So far, only purely computational studies have been used to test the predictive force of Eq. (2) and those tests targeted highly-non-Hermitian systems, i.e., low-Q plasmonic nanoantennas [21] and metallic gratings [22].

Important open questions surround the proposed alternative perturbation formula, Eq. (2). For instance, even if it is evident that strongly non-Hermitian systems like low-Q plasmonics require a revised perturbation theory, one wonders which genuine benefits, if any, can be expected from Eq. (2) for high-$Q$ microcavities since these operate in a manner closely analogous to Hermitian systems with infinitesimal absorption or leakage [23]. More fundamentally, the question arises in QNM theory if the concept of complex mode volume introduced in [19] is just an abstract mathematical construct, or carries true physical significance. In particular, the question of the physics captured by Im $\widetilde{V}$ in Eq. (2) arises, for which simple intuitive arguments have not yet been presented in earlier works [21,22]. Finally, we note that no experiment has validated Eq. (2) so far. Even beyond the question whether this equation correctly captures real perturbation experiments, such an experiment could for the first time test if QNMs, which are widely regarded as difficult mathematical objects with complex frequencies and divergent fields, are in fact directly measurable physical objects that can be mapped through unique signatures in experiments.

This Letter answers all three questions. In particular, we provide experimental evidence that the perturbation theory of high-$Q$ microcavities, like low-$Q$ resonators, should rely on non-Hermitian physics. Second, as a direct consequence of the relation between $\Delta\widetilde{\omega}$ and $\widetilde{V}$ in Eq. (2), we show that our perturbation measurements of $\Delta\widetilde{\omega}$ allow for a direct mapping of the spatial distribution of $\widetilde{V}$. This is an important result since $\widetilde{V}$ is deeply involved in important phenomena of light-matter interactions in non-Hermitian open systems, e.g., Purcell effect, strong coupling [8]. We also conclusively clarify the physics captured by Im $\widetilde{V}$. Finally we provide the first analysis of the validity domain of Eq. (2), pinpointing the physics that causes the breakdown even of revised perturbation theory.

Our main experimental results, obtained for a PhC cavity formed by four missing holes organized in an hexagonal array of holes, are summarized in Fig. 1. Electron beam lithography followed by reactive ion etching is used to fabricate the perforated air-membrane [15]. InAs quantum dots emitting at 1300 nm and excited at 780 nm are embedded in the GaAs membrane. We use a commercial Scanning Near-field Optical Microscope (SNOM) from TwinSNOM-Omicron in illumination/collection configuration. The fiber tip, a chemically etched, uncoated near-field fiber probe [15], plays the role of the perturber and the probe. It is raster scanned at a constant height above the membrane surface, and for each position, we record the fluorescence spectrum, see SM for details. By fitting the recorded lineshape with a Lorentzian profile, we infer the resonance wavelength and the $Q$. Three spectra recorded for three tip positions, labeled A, B and C in Fig. 1a, are plotted in Fig. 1c.

The results, shown with the resonance-shift map in Fig. 1b, are in quantitative agreement with previous reports [14,4,15,24] showing resonance red-shifts with tiny dielectric perturbers. We estimate that the spatial resolution, which defines the dimension of the tip perturbation, is ≈ 70 nm. More important in the present context are the tip-induced variations of $Q$, whose map in Fig. 1d shows both $Q$-increases and $Q$-decreases for the first time [25]. In order to link all these values to the intrinsic cavity $Q$ (without perturber), we additionally repeat the SNOM scans for different tip distances $d$ with respect to the membrane interface. Note that the minimum separation distance, $d_{min} \approx 30$ nm, depends on the tip-interface interaction and cannot be accurately measured. The data recorded for the three tip positions are given in Fig. 1e. The three series of data all tend to $Q = 2300 \pm 40$, which is also the intrinsic $Q$ value measured when the tip is 1-µm away from the sample. An important and simple outcome of Figs. 1c-e is that the *same* perturber may either increase $Q$ (point A), leave $Q$ unchanged (point C) or decrease $Q$ (point B). Therefore our hyperspectral mapping of the of the QNM near field refutes the general validity of Eq. (1). Further analysis of the experimental $\Delta\tilde{\omega}$ map will be provided afterwards.

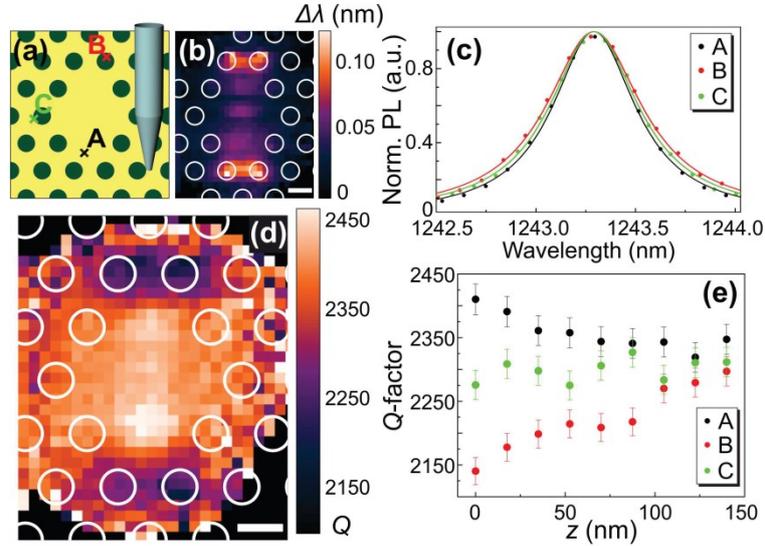

FIG. 1 Experimental results. (**a**) Sketch of the PhC cavity. (**b**) Wavelength-shift map as the tip is scanned over the cavity, with superimposed holes. (**c**) Photoluminescence recorded for 3 tip positions, A, B and C shown in (a). Curves are Lorentzian fits of the data small points. The black and red points are blue-shifted by 0.05 and 0.08 nm to ease the visual comparison for cavity $Q$'s. (**d**) Perturbation-induced $Q$ map. (**e**) $Q$ as a function of the offseted distance $z - d_{min}$ between the tip and the PhC membrane. Conclusively, the same tip may either enhance or decrease the intrinsic $Q = 2300 \pm 40$, depending on its position. The PhC parameters are: lattice period $a = 331$ nm, hole diameter ≈ 206 nm, and GaAs-membrane thickness 320 nm.

To quantitatively test Eq. (2) for high-$Q$ cavities and quantify its domain of validity, we consider the same geometry and material as in the experiment (the membrane refractive index is assumed to be 3.46), and replace the tip by a deep-subwavelength dielectric perturber (volume $V_p$, permittivity $\Delta\varepsilon + \varepsilon_b$ with $\varepsilon_b \equiv \varepsilon(\mathbf{r}_0)$). We compute the resonance mode of the unperturbed cavity with the QNM-solver QNMEig [26] implemented in COMSOL Multiphysics [27]. QNMEig provides normalized QNMs $[\tilde{\mathbf{E}}, \tilde{\mathbf{H}}]$, with $\iiint \left(\varepsilon\tilde{\mathbf{E}}^2 - \mu_0\tilde{\mathbf{H}}^2\right) d^3\mathbf{r} = 1$, and $\tilde{V}(\mathbf{r}_0)$ is simply given by $\left(2\varepsilon(\mathbf{r}_0)\tilde{\mathbf{E}}^2(\mathbf{r}_0)\right)^{-1}$. The computed eigenfrequency is $\tilde{\lambda} = 2\pi c/\tilde{\omega} = 1364 + i0.13$ nm, implying that the computed $Q$ is twice larger than the experimental one, probably because of losses induced by layer absorption, surface roughness or other extrinsic effects. A spatial map of $|\tilde{\mathbf{E}}|^2$ in a plane 30-nm above the cavity surface is shown in Fig. 2(a).

Figure 2(b) compares the $\Delta Q$'s predicted with Eq. (2) with exact values computed by solving the perturbed cavity. The data are obtained for a dipole polarizability $\alpha = 4\pi R^3 \frac{\varepsilon_{SiO_2}/\varepsilon_0 - 1}{\varepsilon_{SiO_2}/\varepsilon_0 + 2}$ corresponding to the static polarizability of a silica ($\varepsilon_{SiO_2} = 2.25\varepsilon_0$) nanosphere in air of radius $R = 55$ nm. Since we use exactly the same mesh for the two computations, numerical dispersion is negligible and the comparison strictly quantifies the error due to the single mode approximation. Figure 2(c) compares the $\Delta Q$ predictions of Eq. (2) with exact numerical values for increasing values of the perturber polarizability $\alpha$, assumed to be real. Three perturber locations, corresponding to the three tip positions used in the experiment, are considered. Remarkably, our key experimental observation that the same perturber may either decrease or increase $Q$ as its position is varied, independently of the wavelength-shift sign, is well captured by Eq. (2).

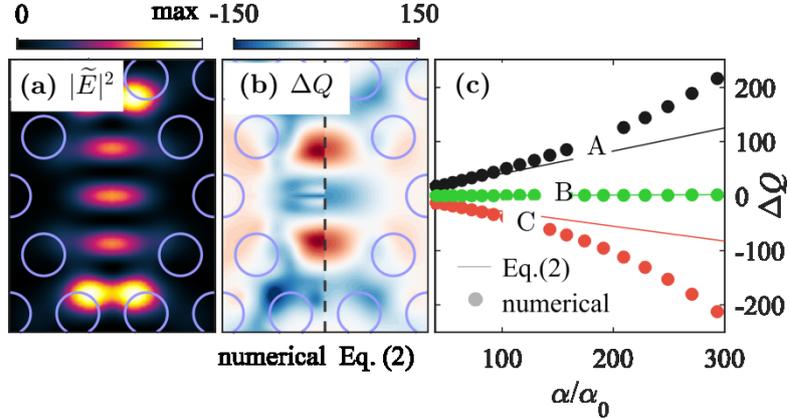

FIG. 2 Numerical test of Eq. (2) for the cavity used in the experiment. (**a**) Maps of $|\tilde{\mathbf{E}}|^2$. (**b**) Comparison between the $\Delta Q$ maps predicted with Eq. (2) (left) and exact values (right) for $\alpha = 166\,\alpha_0$. (**c**) Validity of Eq. (2) for increasing values of the polarizability and for the three tip positions, A, B and C, used in the experiment. $\alpha_0$ denotes the static polarizability of a 10-nm-radius silica sphere in air, so that the full horizontal scale covers silica spheres with radii from 10 to 70 nm. Note that Eq. (1) predicts $\Delta Q = 0$ for all positions and all $\alpha$. In (b) and (c), the point dipole perturber is assumed to be located in a plane 30 nm above the semiconductor PhC membrane, and the exact values are computed by iteratively searching the complex-frequency pole of Eq. (4) for $\mathbf{E}_b = 0$ with the regularized scattering tensor $\Delta\mathbf{G}(\mathbf{r},\mathbf{r}',\omega)$ computed with COMSOL Multiphysics [27].

As expected, Fig. 2(c) evidences that for vanishing $\alpha$'s, Eq. (2) is virtually exact. However, some differences, not observed in previous studies for low-$Q$ plasmonic structures [21-22], are observed for $\alpha > 150\,\alpha_0$. This leads us to the important question of the conditions under which Eq. (2) may be used with confidence and what parameters are impacting its domain of validity. For clarification, let us briefly recall the approximations needed to derive the perturbation formula. We focus on perturbations so small that the point-dipole approximation applies, in which case the perturber acts as an induced dipole moment $\mathbf{p}\delta(\mathbf{r} - \mathbf{r}_0)$. The total incident field driving the dipole is the sum of the external field $\mathbf{E}_b(\mathbf{r}_0, \omega)$ and the field scattered by the cavity onto the dipole

$$\mathbf{p} = \boldsymbol{\alpha}(\omega)\{\mathbf{E}_b + \mu_0\omega^2\Delta\mathbf{G}(\mathbf{r}_0, \mathbf{r}_0, \omega)\mathbf{p}\}, \tag{3}$$

where the polarizability $\boldsymbol{\alpha}(\omega) = \alpha(\omega)\mathbf{1}$ [28] is defined for a perturber placed in an homogenous medium of permittivity of permittivity $\varepsilon(\mathbf{r}_0, \omega)$, $\Delta\mathbf{G}$ is a regularized scattering tensor [29] satisfying $\mathbf{G} = \mathbf{G}_0 + \Delta\mathbf{G}$, with $\mathbf{G}(\mathbf{r}_0, \mathbf{r}, \omega)$ and $\mathbf{G}_0$ the Green tensors of the unperturbed cavity and of the uniform medium of permittivity $\varepsilon(\mathbf{r}_0, \omega)$, respectively. $\Delta\mathbf{G}$ encompasses both the coupling to the unperturbed mode of interest, and to all the other cavity modes, and accordingly, is expressed as [8]

$$\Delta\mathbf{G}(\mathbf{r},\mathbf{r}',\omega) = \frac{-\widetilde{\omega}\tilde{\mathbf{E}}_N(\mathbf{r})\otimes\tilde{\mathbf{E}}_N(\mathbf{r}')}{\mu_0\omega^2(\omega-\widetilde{\omega})} + \delta\mathbf{G}(\mathbf{r},\mathbf{r}',\omega), \tag{4}$$

where the first term represents the contribution from the non-degenerated (and normalized, see below) mode $\tilde{\mathbf{E}}_N(\mathbf{r})$ of interest, while the second term gathers the contribution of all other cavity modes and a continuum of radiation modes for cavities located in non-uniform backgrounds, on substrates for instance [26]. Full analyticity is recovered by neglecting the $\delta\mathbf{G}$ term in Eq. (4). Doing so and injecting Eq. (3) into Eq. (4) in the absence of the external field $\mathbf{E}_b(\mathbf{r}_0, \omega)$, we directly obtain Eq. (2).

In the SM, we analyze the impact of omitting $\delta \mathbf{G}$. Since both terms in the right-hand side of Eq. (4) depend on the perturber position differently, we have to make several approximations. We assume real values for the polarizability α and neglect the vectorial character of the coupling, approximating $\delta \mathbf{G}$ by a diagonal tensor with identical diagonal terms $\delta G$ equal to one third of the trace of $\delta \mathbf{G}$. This way, we find two upper bounds for α to obtain accurate predictions of $\Delta \widetilde{\omega}$ with Eq. (2). For Re $\Delta \widetilde{\omega}$, $\alpha < \alpha_r \equiv \left|\frac{1}{\mu_0 \operatorname{Re}(\widetilde{\omega}^2 \delta G)}\right|$, and for Im $\Delta \widetilde{\omega}$, two conditions have to be satisfied $\alpha < \alpha_r$ and $\alpha < \alpha_i \equiv \left|\frac{1}{\mu_0 \operatorname{Im}(\widetilde{\omega}^2 \delta G)}\right| \left|\frac{\operatorname{Im} \widetilde{V}^{-1}}{\operatorname{Re} \widetilde{V}^{-1}}\right|$, implying that Im($\Delta \widetilde{\omega}$), i.e. $\Delta Q$, can be predicted, at best, with the same accuracy as $\Delta \lambda$, but not better. Moreover, as $Q$ increases, $\left|\frac{\operatorname{Im} \widetilde{V}^{-1}}{\operatorname{Re} \widetilde{V}^{-1}}\right|$ decreases towards zero, and so does $\alpha_i$. Therefore, it is more difficult to predict $\Delta Q$ accurately for a high-$Q$ cavity than for a low-$Q$ one. This explains why no visible deviation between the predictions of Eq. (2) and exact numerical data have been detected in earlier works on plasmonic nanoresonators [21-22]. Finally, for our present cavity, strong near-field interactions between the perturber and the PhC membrane result in $|\operatorname{Re} \delta G| \gg |\operatorname{Im} \delta G|$ for all perturber positions (see SM for specific numerical values). This explains why the predictions of $\Delta Q$ in Fig. 2(c) and those of $\Delta \lambda$ in Fig. SI-4 in SM are equally accurate over the entire range of polarizability values.

The success of Eq. 2 to predict $Q$-changes resides in the replacement of a real mode volume by a complex one, and more precisely, of $\left|\widetilde{\mathbf{E}}(\mathbf{r}_0)\right|^2$ by $\widetilde{\mathbf{E}}^2(\mathbf{r}_0)$ in the denominator of $\widetilde{V}$. This replacement preserves the phase information $\phi(\mathbf{r}_0)$ of the mode at the perturber location. For an intuitive picture that explains why the phase is essential, consider a driving field impinging onto a perturbed cavity. The field does not see the tiny perturber and in first instance excites the cavity as if it were unperturbed. The cavity then directly scatters in free space and also excites the perturber, which in turn re-excites the cavity mode with a round-trip dephasing delay of $2\phi(\mathbf{r}_0)$. The total radiated field by the cavity results from the interference of the direct initial radiation and the delayed one. Depending on whether these interferences are constructive or destructive, the total cavity radiation can be higher or lower than the intrinsic cavity radiation, possibly allowing for either an increase or a decrease of $Q$. This *a posteriori* explains why Eq. (1) that relies on an $\widetilde{\mathbf{E}} \cdot \widetilde{\mathbf{E}}^*$ product and hence loses the phase information, fails to predict $Q$-changes.

The concept of complex $\widetilde{V}'s$ is recent [19]. It seems to be rooted in important phenomena of light-matter interactions in non-Hermitian open systems [8]. For instance, the ratio $\operatorname{Im} \widetilde{V}^{-1}/\operatorname{Re} \widetilde{V}^{-1}$ quantifies the spectral asymmetry of the mode contribution to the modification of the spontaneous emission rate of an emitter weakly coupled to a cavity [19]. For strong coupling, it modifies the usual expression of the Rabi frequency [9] by blurring and moving the boundary between the weak and strong coupling regimes [8,30]. Despite these strong roots, complex $\widetilde{V}'s$ are often seen as a mathematical abstraction. In fact, Eq. 2 and our experiment show that complex $\widetilde{V}'s$ are not just a mathematical tool, but in fact are directly measurable.

Figure 3 shows the maps of Re $\widetilde{V}^{-1}$ and Im $\widetilde{V}^{-1}$, which have been directly inferred from our $\Delta \widetilde{\omega}$ measurements by injecting a tip polarizability $\alpha_{tip} = 166 \, \alpha_0$ (tip curvature radius of $R = 55$ nm) in Eq. (2). For comparison, we also plot the theoretical maps computed with the QNM-solver. Note that to allow for a better comparison, we have multiplied the experimental values of Re $\widetilde{V}^{-1}$ and Im $\widetilde{V}^{-1}$ by a $\times 1/4$ rescaling factor. The latter corresponds to a tip radius only 30% larger ($R = 73$ nm), and can be understood by considering that a static sphere dipolar polarizability is a simplistic model for the tip used in our experiment. There are differences between the experimental maps and the computed ones. Nevertheless, the experimental and theoretical maps qualitatively share the same dominant features, notably a successful agreement on the *locations* and *amplitudes* of the minimum and maximum values, and an overall 10-fold difference between Re $\widetilde{V}^{-1}$ and Im $\widetilde{V}^{-1}$.

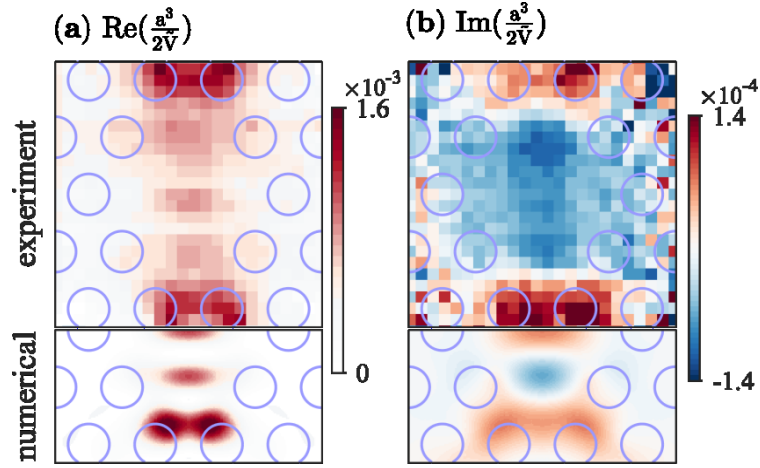

FIG. 3 Maps of (**a**) $\text{Re}(\tilde{V}^{-1})$ and (**b**) $\text{Im}(\tilde{V}^{-1})$ computed with the QNM-solver 30 nm above the semiconductor membrane (bottom) and directly inferred from the $\Delta\tilde{\omega}$ measurements using Eq. (2) with a tip polarizability $\alpha_{tip} = 166\ \alpha_0$ (top). Note that the experimental values are all rescaled by a factor ¼.

To summarize, we have demonstrated, with hyperspectral-imaging near-field experiments, that the perturbation theory of high-$Q$ microcavities should rely on complex modal volumes to fully account for the role of the perturber at the nanoscale. This demonstration is a first and direct evidence of the effects of complex modal volumes, arising from the intrinsic property of all photonic resonators of being an open (i.e. non Hermitian) system, on the optical response of a photonic system. We have shown that QNM theory allows for a quantitative prediction of both $\text{Re}(\Delta\tilde{\omega})$ and $\text{Im}(\Delta\tilde{\omega})$ as a function of the perturber position, whereas classical theory based on Hermitian physics only gives access to $\text{Re}(\Delta\tilde{\omega})$. Equation (2) combines great simplicity and predictive power. It may find applications in various problems related to sensing or trapping, as the additional information provided by dual maps may help lifting the degeneracy of single $\Delta\lambda$-maps, for instance allowing not only the detection of binding event in sensing but also the binding location [11]. Other perspectives concern the analysis of the impact of fabrication imperfections on $Q$'s, post-fabrication $Q$-control [12], optimization of cavities with large $Q$'s, or inverse design of cavities with tailored $\Delta\lambda$ and $\Delta Q$-maps. Equation (2) also offers the possibility to perform direct measurement of the complex mode volume of microcavities, giving greater visibility and operational capacity to an important physical quantity of resonant light-matter interactions.

## Acknowledgements

The authors deeply thank M. Petruzzella, F. W. M. van Otten and A. Fiore of Eindhoven University of Technology for the growth and fabrication process of the sample. W.Y. acknowledges a fellowship of the Centre National de la Recherche Scientifique (CNRS). The work was supported by the French National Agency for Research (ANR) under the project "Resonance" (Grant No. ANR-16-CE24-0013). This study has been carried out with financial support from the French State, managed by the French National Agency for Research (ANR) in the frame of the "Investments for the Future" Programme IdEx Bordeaux – LAPHIA (Grant No. ANR-10-IDEX-03-02). P.L. is pleased to acknowledge the support from the LabEx LAPHIA and CNRS. This work is part of the research programme of the Netherlands Organisation for Scientific Research (NWO) and was performed at the research institute AMOLF. AFK gratefully acknowledges an NWO-Vici grant for financial support.

## References

[1] H. A. Bethe and J. Schwinger, N.D.R.C. Rpt. D1-117 Cornell University, (March, 1943).
"Perturbation theory for cavities"
[2] R. A. Waldron, Proc. Inst. Electr. Eng. **107C**, 272 (1960).
"Perturbation theory of resonant cavities"

[3] O. Klein, D. M. Dressel, G. Grüner, Int. J. Infrared Milli. Waves **14**, 2423-57 (1993).
"Microwave cavity perturbation techniques: Part I: Principles"
[4] L. Lalouat, B. Cluzel, P. Velha, E. Picard, D. Peyrade, J.P. Hugonin, P. Lalanne, E. Hadji, F. De Fornel, Phys. Rev. B **76**, 041102 (2007).
"Near-field interactions between a subwavelength tip and a small-volume photonic-crystal nanocavity"
[5] L. Ramunno and S. Hughes, Phys. Rev. B **79**, 161303(R) (2009).
"Disorder-induced resonance shifts in high-index-contrast photonic crystal nanocavities"
[6] M. Burresi, T. Kampfrath, D. van Oosten, J. C. Prangsma, B. S. Song, S. Noda, and L. Kuipers, Phys. Rev. Lett. **105**, 123901 (2010).
"Magnetic Light-Matter Interactions in a Photonic Crystal Nanocavity"
[7] L. Neumeier, R. Quidant and D.E. Chang, New J. Phys. **17**, 123008 (2015).
"Self-induced back-action optical trapping in nanophotonic systems"
[8] P. Lalanne, W. Yan, K. Vynck, C. Sauvan, and J.-P. Hugonin, Laser Photonics Rev. **12**, 1700113 (2018).
"Light interaction with photonic and plasmonic resonances"
[9] J.-M. Gérard, Top. Appl. Phys. **90**, 269 (2003).
"Solid-state cavity-quantum electrodynamics with self-assembled quantum dots"
[10] F. Vollmer and S. Arnold, Nature Methods **5**, 591-596 (2008).
"Whispering-gallery-mode biosensing: label-free detection down to single molecules"
[11] K. D. Heylman, K. A. Knapper, E. H. Horak, M. T. Rea, S. K. Vanga, and R. H. Goldsmith, Adv. Mater. **29**, 1700037 (2017).
"Optical Microresonators for Sensing and Transduction: A Materials Perspective"
[12] A. Badolato, K. Hennessy, M. Atatüre, J. Dreiser, E. Hu, P.M. Petroff, A. Imamoğlu, Science **308**, 1158-1161 (2005).
"Deterministic Coupling of Single Quantum Dots to Single Nanocavity Modes"
[13] A.F. Koenderink, M. Kafesaki, B.C. Buchler, V. Sandoghdar, Phys. Rev. Lett. **95**, 153904 (2005).
"Controlling the Resonance of a Photonic Crystal Microcavity by a Near-Field Probe"
[14] S. Mujumdar, A. F. Koenderink, T. Sünner, B. C. Buchler, M. Kamp, A. Forchel, and V. Sandoghdar, Opt. Express **15**, 17214–20 (2007).
"Near-field imaging and frequency tuning of a high-Q photonic crystal membrane microcavity"
[15] F. Intonti, S. Vignolini, F. Riboli, A. Vinattieri, D. S. Wiersma, M. Colocci, L. Balet, C. Monat, C. Zinoni, L. H. Li, R. Houdré, M. Francardi, A. Gerardino, A. Fiore, and M. Gurioli, Phys. Rev. B **78**, 041401(R) (2008).
"Spectral tuning and near-field imaging of photonic crystal microcavities"
[16] N. Le Thomas and R. Houdré, Phys. Rev. B **84**, 035320 (2011).
"Inhibited emission of electromagnetic modes confined in subwavelength cavities"
[17] F. Ruesink, H.M. Doeleman, R. Hendrikx, A. F. Koenderink, E. Verhagen, Phys. Rev. Lett. **115**, 203904 (2015)
"Perturbing Open Cavities: Anomalous Resonance Frequency Shifts in a Hybrid Cavity-Nanoantenna System"
[18] S. Vignolini, F. Intonti, F. Riboli, L. Balet, L.H. Li, M. Francardi, A. Gerardino, A. Fiore, D.S. Wiersma, and M. Gurioli, Phys. Rev. Lett. **105**, 123902 (2010).
"Magnetic Imaging in Photonic Crystal Microcavities"
[19] C. Sauvan, J.P. Hugonin, I.S. Maksymov and P. Lalanne, Phys. Rev. Lett **110**, 237401 (2013).
"Theory of the spontaneous optical emission of nanosize photonic and plasmon resonators"
[20] E. A. Muljarov and W. Langbein, Phys. Rev. B **93**, 075417 (2016).
"Resonant-state expansion of dispersive optical open systems: creating gold from sand"
[21] J. Yang, H. Giessen, P. Lalanne, Nano Lett. **15**, 3439 (2015).
"Simple Analytical Expression for the Peak-Frequency Shifts of Plasmonic Resonances for Sensing"
[22] T. Weiss, M. Mesch, M. Schäferling, and H. Giessen, W. Langbein and E. A. Muljarov, Phys. Rev. Lett. **116**, 237401 (2016).

"From Dark to Bright: First-Order Perturbation Theory with Analytical Mode Normalization for Plasmonic Nanoantenna Arrays Applied to Refractive Index Sensing"

[23] A. Oskooi and S. G. Johnson, *Electromagnetic Wave Source Conditions*, Chapter 4 in Advances in *FDTD Computational Electrodynamics: Photonics and Nanotechnology*, A. Taflove, A. Oskooi, and S. G. Johnson, eds., Norwood, MA: Artech House, 2013.

[24] N. Caselli et al., Light: Science & Applications **4**, e326 (2015).
"Ultra-subwavelength phase-sensitive Fano-imaging of localized photonic modes"

[25] $Q$-increases by modifying cavity geometry have been previously reported using slabs [16] and scatterer gratings [17] in near fields, but not with a localized perturbation, nor with scanning through the mode to determine the relation between $Q$-changes and mode distributions. Perturbations by extended structures like slabs, have rather been understood as radiation pattern engineering to control $Q$.

[26] W. Yan, R. Faggiani, P. Lalanne, "Rigorous modal analysis of plasmonic nanoresonators", Phys. Rev. B **97**, 205422 (2018). QNMEig and companion Matlab Toolboxes are available at the last author group webpage.

[27] https://www.comsol.com/ (Version 5.2a).

[28] The present theory is derived for isotropic and non-magnetic perturbers for the sake of simplicity, but these assumptions can be easily removed.

[29] L. Novotny and B. Hecht, *Principles of Nano-Optics*, Chapt. 15 (Cambridge University Press, Cambridge, 2006).

[30] E. Lassale, N. Bonod, T. Durt and B. Stout, Opt. Lett. **43**, 1950 (2018).
"Interplay between spontaneous decay rates and Lamb shifts in open photonic systems"

# Mapping Complex Mode Volumes with Cavity Perturbation Theory

K. G. Cognée[1,2], W. Yan[1], F. La China[3], D. Balestri[3], F. Intonti[3], M. Gurioli[3], A. F. Koenderink[2], P. Lalanne[1]*

[1]LP2N, Institut d'Optique, CNRS, Univ. Bordeaux, Talence, 33400, France
Center for Nanophotonics, AMOLF, Science Park 104, 1098XG, Amsterdam The Netherlands
[3]LENS, University of Florence, Sesto Fiorentino, 50019, Italy
E-mail: philippe.lalanne@institutoptique.fr

This supplementary Material provides complementary results and discussions to the main text, starting with a Section detailing discussing the reliability of the $\Delta Q$-measurements, followed by a formal comparison of the classical perturbation formula of Eqs. (1) and (2), a study of the accuracy of Eq. (2) for predicting resonance shifts, and by an analytical study of the domain of validity of Eq. (2) that leads to upper bounds for the maximum perturber strength.

## 1. Experimental details

We use the tip to excite the embedded InAs QD with a cw laser at 780 nm and, for every tip position, we measure the QD photoluminescence spectrum. At room temperature, the spectrum covers more than 100 nm. It exhibits a Lorentzian peak for each cavity resonance. Due to the interaction with the tip, $\text{Re}(\widetilde{\omega})$ and $\text{Im}(\widetilde{\omega})$ are both modified. By fitting the spectra for every tip position, we obtain the $\Delta\lambda$ and $Q$ maps reported in Fig. 1.

The feedback mechanism of the SNOM is able to maintain the tip on the sample surface at constant height, whenever the sample is flat. In photonic crystal cavities, it forces the tip to follow the topography and then, when the tip is on an air pore, the tip height is reduced by few tens of nm. The measure of the tip-height map, see Fig. S1a, allows us to reconstruct, *a posteriori*, the perturbation map with a spatial alignment of a few tens of nanometers, which is needed for a comparison with theoretical prediction. The $z$-scan is then performed by moving the sample vertically with steps of 20 nm. During the vertical scan, the tip is maintained at a constant height. Then, after each $z$-scan, the sample is repositioned thank to the feedback mechanism to keep the spatial alignment, and then is moved in the $(x, y)$ plane.

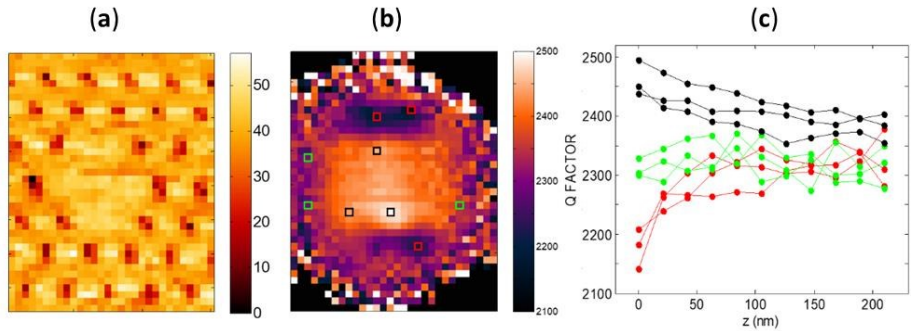

Fig. SI-1 (**a**) Topography map. (**b**) Corresponding $Q$-variation map. The colored squares represent several tip locations. (**c**) $Q$-variation as a function of the offseted distance $z - d_{min}$ between the tip and the photonic-crystal membrane. The black, green and red curves are obtained for tip locations shown in (c) with the squares of the same colors.

In order to detect possible systematic errors in the $z$-scan (such as sample/tip drift), we repeated the $z$-scan several times for different $(x, y)$ locations. An example is shown in Figs. SI1b-c. Figure SI1b reports the $Q$ map at $z = 0$ for three A points (red squares), three B points (black squares) and three C points (green squares), all located at quite different position. Figure SI1c reports the $Q$-variation for every points. All the data converge to a common value with similar trends, denoting the reliability of the presented data.

Finally we address the repeatability of the SNOM measurements to detect possible artefacts. Figure SI-2 shows three different maps of the $Q$-factor obtained with the same tip during three different days. The data comparison conclusively evidences a quantitative agreement between the three sets of data.

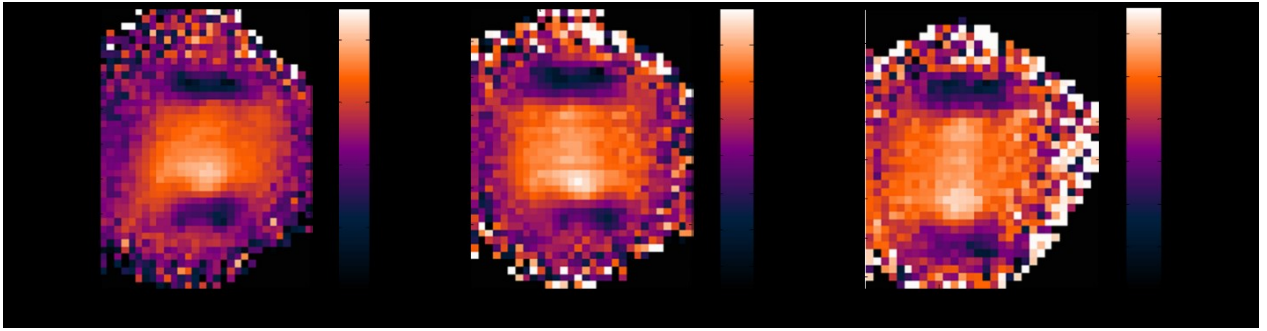

Fig. SI-2 Three different maps of the $Q$-variation induced by the SNOM tip measured for three different days without changing the tip.

## 2. Formal comparison of Eqs. (1) and (2)

The main difference between Eqs. (1) and (2) resides in the replacement of $\tilde{\mathbf{E}} \cdot \tilde{\mathbf{E}}^*$ product by the unconjugated product $\tilde{\mathbf{E}} \cdot \tilde{\mathbf{E}}$. In order to clarify the impact of the replacement for high-$Q$ microcavities, we consider perturbations formed by deep-subwavelength isotropic dielectric perturbers (volume $V_p$, permittivity $\Delta\varepsilon + \varepsilon_b$) that are introduced into a background material of permittivity $\varepsilon_b$. In the static limit, the perturbers act as point isotropic electric dipoles, with a polarizability proportional to their perturber volume, $\alpha = \alpha' V_p$, $\alpha'$ being a dimensionless coefficient. For spherical perturbers at optical frequencies, $\alpha' = \frac{3\Delta\varepsilon}{\Delta\varepsilon + 3\varepsilon_b}$ with $-1.5 < \alpha' < 3$ for perturbers with a positive permeability. Replacing $\alpha$ in Eq. (2), and assuming that $\alpha'$ is a real number, we get

$$\frac{\text{Re}(\Delta\tilde{\omega})}{\text{Re}(\tilde{\omega})} \approx -\frac{\alpha'}{2}\left[\text{Re}\left(\frac{V_p}{\tilde{V}}\right) - \frac{1}{2Q}\text{Im}\left(\frac{V_p}{\tilde{V}}\right)\right], \tag{SI-1a}$$

$$\frac{\text{Im}(\Delta\tilde{\omega})}{\text{Im}(\tilde{\omega})} \approx -\frac{\alpha'}{2}\left[\text{Re}\left(\frac{V_p}{\tilde{V}}\right) + 2Q\,\text{Im}\left(\frac{V_p}{\tilde{V}}\right)\right]. \tag{SI-1b}$$

For high-$Q$ photonic cavities, $\text{Im}(\tilde{\mathbf{E}}) \ll \text{Re}(\tilde{\mathbf{E}})$ and $\text{Re}\left(\frac{V_p}{\tilde{V}}\right) \approx \frac{V_p}{V} \gg \frac{1}{2Q}\text{Im}\left(\frac{V_p}{\tilde{V}}\right)$ (remember that $V$ is the approximate mode volume defined in Eq. (1)), so that Eq. (SI-1a) reduces to

$$\text{Re}(\Delta\tilde{\omega}) \approx -\omega_0 \alpha' \frac{V_p}{V}, \tag{SI-2a}$$

which is exactly the shift predicted by Eq. (1). Note that this conclusion does not hold for low-$Q$ plasmonic resonators. Quite the contrary, the imaginary part $\text{Im}\left(\frac{V_p}{\tilde{V}}\right)$ cannot be neglected in general in Eq. (SI-1b). Even for our microcavity, Fig. 2a evidences that the imaginary part dominates over the real part $\frac{1}{2Q}\text{Re}\left(\frac{V_p}{\tilde{V}}\right)$, i.e., $\text{Im}\left(\frac{V_p}{\tilde{V}}\right) \gg \frac{1}{2Q}\text{Re}\left(\frac{V_p}{\tilde{V}}\right)$ and

$$\text{Im}(\Delta\tilde{\omega}) \approx -\frac{\alpha'}{2}\text{Re}(\tilde{\omega})\,\text{Im}\left(\frac{V_p}{\tilde{V}}\right). \tag{SI-2b}$$

This mathematically justifies why Eq. (1) fails at predicting $\text{Im}(\Delta\tilde{\omega})$.

## 3. Accuracy of Eq. (2) to predict resonance shifts

The main text focuses on the prediction of perturbation-induced quality-factor changes, $\Delta Q$, which is the novelty of the work. For the sake of completeness, we have performed a similar study for the resonance shift, whose main result are summarized in Figs. SI-3 and SI-4. The conclusion for our cavity geometry is that Eq. (2) is as accurate to predict wavelength shifts, as it is at predicting $Q$-changes. We believe that this result represents a strong evidence of the great added value brought by Eq. (2) for high-$Q$ cavities.

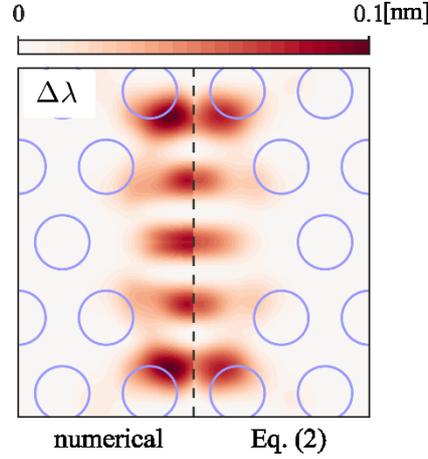

Fig. SI-3 Validation of Eq. (2) for the resonance wavelength shift $\Delta\lambda$ by comparison with exact numerical data obtained by solving the perturbed cavity. All simulations are obtained with the same structure and perturber as in the Fig. 2 in the main text.

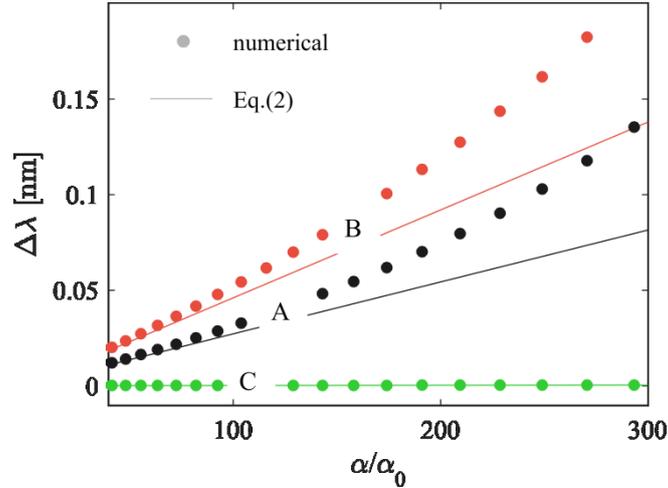

FIG. SI-4 Study of the validity range of Eq. (2) for predicting resonance wavelength shifts $\Delta\lambda$ by comparison with fully-numerical data obtained by solving the perturbed cavity. The perturber polarizability $\alpha$ is normalized by $\alpha_0$, the static polarizability of a silica sphere with 10-nm radius in air. Three perturber positions, 30 nm above the semiconductor membrane, are considered; they are labelled as "A", "B", "C", corresponding to the same position in Fig. 1(e) in the main text.

## 4. Analytical derivation of the domain of validity of Eq. (2)

As evidence by Fig. 2c in the main text, Eq. (2) is exact in the limit of infinitely small perturbations. In this Section, we would like to quantify under which condition Eq. (2) may be approximately valid and used with confidence to predict both resonance shifts and $\Delta Q$-changes.

To obtain a qualitative insight into the domain of validity of Eq. (2), we start from the *exact* Eqs. (4) and (5), insert Eq. (5) into Eq. (4), annul the driving field ($\mathbf{E}_b = 0$), then perform a Taylor expansion with respect to $\alpha$ up to the second order, and obtain

$$\frac{\Delta\widetilde{\omega}}{\widetilde{\omega}} \approx -\alpha \widetilde{\mathbf{E}}_N \cdot [\mathbf{1} - \alpha\mu_0\omega^2\delta\mathbf{G}] \cdot \widetilde{\mathbf{E}}_N. \tag{SI-2}$$

The second term inside the bracket, $\alpha\mu_0\omega^2\delta\mathbf{G}$ gathers the contribution of all other modes that contribute to the mode density at the cavity, except for the cavity mode that is singled out by $\widetilde{\mathbf{E}}_N$. In the limit that this contribution is negligible, Eq. (SI-2) simply reduces to Eq. (2) in the main text. Accordingly, the validity of Eq. (2) requires that

$$\| \alpha\mu_0\omega^2\delta\mathbf{G}\|_\infty \ll 1, \tag{SI-3}$$

where the operation $\|\cdots\|_\infty$ represents the infinite norm of a matrix.

Though Eq. (SI-3) formally quantifies the domain of validity of Eq. (2), it is difficult to extract more information, since $\delta\mathbf{G}$ is a $3\times 3$ symmetric matrix containing 6 different components. To bypass this difficulty, we make the approximation $\delta\boldsymbol{G} \approx \delta G\,\mathbf{1}$, with $\delta G \equiv \text{Tr}(\delta\boldsymbol{G})/3$, i.e., neglecting the vectorial character of $\delta\boldsymbol{G}$, where $\mathbf{1}$ represents the identity matrix, and further assume $\alpha$ is real, i.e., neglecting radiation loss and material absorption of the perturber. Under these approximations, we compare $\Delta\widetilde{\omega}$ predicted from Eq. (SI-2) and Eq. (2), and derive that the dominant conditions for Eq. (2) to accurately predict $\text{Re}(\Delta\widetilde{\omega})$ and $\text{Im}(\Delta\widetilde{\omega})$ are respectively

$$|\alpha| \ll \alpha_r \text{ and } |\alpha| \ll \alpha_i, \tag{SI-4}$$

where $\alpha_r$ and $\alpha_i$ are given by

$$\alpha_r = \min\left\{\left|\frac{1}{\mu_0\,\text{Re}(\omega^2\delta G)}\right|,\, \left|\frac{1}{\mu_0\,\text{Im}(\omega^2\delta G)}\right|\left|\frac{\text{Re}(\widetilde{V}^{-1})}{\text{Im}(\widetilde{V}^{-1})}\right|\right\}, \tag{SI-5a}$$

$$\alpha_i = \min\left\{\left|\frac{1}{\mu_0\,\text{Re}(\omega^2\delta G)}\right|,\, \left|\frac{1}{\mu_0\,\text{Im}(\omega^2\delta G)}\right|\left|\frac{\text{Im}(\widetilde{V}^{-1})}{\text{Re}(\widetilde{V}^{-1})}\right|\right\}. \tag{SI-5b}$$

Note that, to derive Eqs. (SI-4)- (SI-5), we have used the relations, $\text{Re}(\widetilde{V}^{-1}) \gg \frac{1}{2Q}\text{Im}(\widetilde{V}^{-1})$ and $\text{Im}(\widetilde{V}^{-1}) \gg \frac{1}{2Q}\text{Re}(\widetilde{V}^{-1})$ which are valid for high-$Q$ cavities.

The expressions of $\alpha_r$ and $\alpha_i$ can be further simplified by first noting that (1) $\left|\frac{\text{Re}(\widetilde{V}^{-1})}{\text{Im}(\widetilde{V}^{-1})}\right| \gg 1$ for high-$Q$ cavity and (2) we generally have $|\text{Re}(\omega^2\delta G)| \gg \text{Im}(\omega^2\delta G)|$ (as confirmed by numerical simulations[1]) for perturbers placed in the near-field of the cavity. We finally obtain simplified expressions for $\alpha_r$ and $\alpha_i$

$$\alpha_r = \left|\frac{1}{\mu_0\,\text{Re}(\omega^2\delta G)}\right|, \tag{SI-6a}$$

$$\alpha_i = \min\left\{\alpha_r,\, \left|\frac{1}{\mu_0\,\text{Im}(\omega^2\delta G)}\right|\left|\frac{\text{Im}(\widetilde{V}^{-1})}{\text{Re}(\widetilde{V}^{-1})}\right|\right\}. \tag{SI-6b}$$

As a numerical example, we consider the two perturber positions *A* and *C* in Fig. 1a for which noticeable $\Delta Q$ and $\Delta\lambda$ changes are observed. We numerically find that $\alpha_r = \alpha_i = 665\alpha_0$ at position *A*, and $\alpha_r = 702\alpha_0$ at position *B* and $\alpha_i = 541\alpha_0$ at position *C*, where $\alpha_0$ denotes the static polarizability of a silica sphere with 10-nm radius in air like in the main text. For both cases, $\alpha_r$ and $\alpha_i$ have similar values, and this *a posteriori* explains why Eq. (2) is as accurate to predict wavelength shifts, as it is at predicting $Q$-changes.

---

[1] $\delta\mathbf{G}$, and then $\delta G$, has been computed with COMSOL Multiphysics. A reasonable estimate for the typical magnitude of $\delta G$ is that it is essentially the non-resonant contribution to the full system Green function [scattered part strictly] on top of which the resonant cavity mode adds. The non-resonant background is of the same order as the Green function of free space for a perturber placed outside the cavity.